%% file: 00paper.tex
\documentclass[sigconf,authorversion]{acmart}

\usepackage[normalem]{ulem}

\AtBeginDocument{%
  \providecommand\BibTeX{{%
    \normalfont B\kern-0.5em{\scshape i\kern-0.25em b}\kern-0.8em\TeX}}}

\copyrightyear{2021} 
\acmYear{2021} 
\setcopyright{rightsretained} 
\acmConference[SIGIR '21]{Proceedings of the 44th International ACM SIGIR Conference on Research and Development in Information Retrieval}{July 11--15, 2021}{Virtual Event, Canada}
\acmBooktitle{Proceedings of the 44th International ACM SIGIR Conference on Research and Development in Information Retrieval (SIGIR '21), July 11--15, 2021, Virtual Event, Canada}\acmDOI{10.1145/3404835.3462893}
\acmISBN{978-1-4503-8037-9/21/07}

\usepackage{booktabs} 
\usepackage{pgfplots}
\usepackage{bbm}
\usepackage{color}
\usepackage{amsmath}
\usepackage{xfrac}

\usepackage{bbold}  
\usepackage{graphicx}
\usepackage{subcaption}

\newtheorem*{definition*}{Definition}

\usepackage{enumitem}  

\begin{document}

\fancyhead{}
\title{On Interpretation and Measurement of Soft Attributes for Recommendation} 

\author{Krisztian Balog}\authornote{Work done while at Google.}
\affiliation{%
  \institution{University of Stavanger}
  \city{Stavanger}
  \country{Norway}
}
\email{krisztian.balog@uis.no}

\author{Filip Radlinski}
\affiliation{%
  \institution{Google}
  \city{London, UK}
}
\email{filiprad@google.com}

\author{Alexandros Karatzoglou}
\affiliation{%
  \institution{Google}
  \city{London, UK}
}
\email{alexkz@google.com}

\begin{abstract}
We address how to robustly interpret natural language refinements (or critiques) in recommender systems.  In particular, in human-human recommendation settings people frequently use \emph{soft attributes} to express preferences about items, including concepts like the originality of a movie plot, the noisiness of a venue, or the complexity of a recipe. 
While binary tagging is extensively studied in the context of recommender systems, soft attributes often involve subjective and contextual aspects, which cannot be captured reliably in this way, nor be represented as objective binary truth in a knowledge base.  This also adds important considerations when \emph{measuring} soft attribute ranking. We propose a more natural representation as personalized relative statements, rather than as absolute item properties. We present novel data collection techniques and evaluation approaches, and a new public dataset. We also propose a set of scoring approaches, from unsupervised to weakly supervised to fully supervised, as a step towards interpreting and acting upon soft attribute based critiques.
\end{abstract}

\ccsdesc[500]{Information systems~Recommender systems}

\keywords{Soft attributes; Recommendation critiquing; Preference feedback}

\maketitle

\input{01-intro}
\input{02-related}
\input{03-testcoll}
\input{04-analysis}
\input{05-methods}
\input{06-results}
\input{07-conclusions}

\bibliographystyle{ACM-Reference-Format}
\bibliography{references}

\end{document}

%% file: 01-intro.tex
\section{Introduction}

Recommender systems provide personalized recommendations of products or services to users, based on their preferences.  A key element of recommender systems is the ability for users to provide feedback on their likes and dislikes.  This is inherently a sequential interactive process, and there is a growing interest in designing automated systems that solicit users' preferences and provide personalized recommendations via multi-turn dialogue~\citep{Culpepper:2018:RFI,Zhang:2018:TCS,Li:2018:TDC,Radlinski:2017:TFC,Sun:2018:CRS,Christakopoulou:2016:TCR}.
Ideally, such systems should be able to incorporate any natural language user feedback on items~\citep{Radlinski:2017:TFC}.

Imagine a scenario where a conversational recommender has suggested some item.  The user may respond ``show me more like this but...''~\citep{McCarthy:2010:ECR}, explaining desired differences with respect to this \emph{anchor item}.  For example, they might ask for a ``less violent,'' ``more thought-provoking,'' or ``less predictable'' movie.  In case of restaurants, it might be for places that are ``more down home.'' 
While soft attributes are common in natural dialogue~\citep{Radlinski:2019:CCP}, interpreting them (that is, determining the degree to which they apply to items) represents the open challenge that we tackle. In contrast, most past work on attribute-based critiquing has been trained and evaluated with datasets focused on binary and unambiguous attributes (e.g., movie tags \cite{Harper:2015:MDH} or book and product categories~\cite{GoodReads:2018,AmazonBooks:2019}).

The key difficulty stems from the unconstrained natural language of soft attributes, i.e.,~properties that are not universally agreed upon, or where different people have different norms, expectations, and thresholds (e.g., ``exciting'' vs. ``boring'' movies). This contrasts with well-studied objective properties like movie actors and genres, point-of-interest categories, or attributes like ``plot twist.''
Due to their subjective nature, representing soft attributes as property-value pairs in a knowledge base is error-prone and often infeasible.
Critically, while soft attributes resemble social tags, there are at least two key differences that make applying tag collections (e.g., the MovieLens Tag Genome~\cite{TagGenome2012}) challenging.
First, tags are binary labels that do not allow for a relative ordering within relevant items (e.g., which of two movies is ``more violent,'' if both are violent to some degree). Thus most tag-based evaluations focus on comparing items with and without a tag, being insensitive to fine-grained structure. 
Second, while in principle both tags and soft attributes are unconstrained natural language, most tagging approaches intentionally bias users towards a consistent vocabulary, for example displaying frequent past labels \cite{Vig:2010:MovieLensTags,Harper:2015:MDH} or pre-defining an ontology of genres \cite{GoodReads:2018}. Soft attributes are by definition meant to allow critiquing in natural conversations, i.e., serve a different purpose.  

Specifically, this paper addresses the problem of the \emph{interpretation} of soft attributes. 
Note that this is different from the task of incorporating subjective item descriptions into latent user preference representations for improving end-to-end recommendation performance (measured in terms of success rate or the number of conversational turns)~\citep{Wu:2019:DLB,Luo:2020:LLC,Luo:2020:DCV,Li:2020:ROA}.
Instead, our objectives are to be able to (1) explicitly measure the degree to which a soft attribute applies to a given item and (2) quantify the degree of ``softness'' (i.e., subjectivity) of soft attributes.

As our first main contribution, we develop a reusable test collection, comprising a set of soft attributes and ground truth item orderings with respect to that attribute (for particular users), and an evaluation metric.
The soft attributes we study come from realistic conversations of movie preferences~\citep{Radlinski:2019:CCP}.  Sampling attributes mentioned extemporaneously by participants, we develop a novel controlled multi-stage crowd labeling mechanism to collect ground truth of personalized partial orderings, while keeping workers' cognitive load low. Contrasting the attributes with tags, we note that only 47\% of the soft attributes we study are present as tags in the commonly used MovieLens TagGenome~\cite{TagGenome2012}.
To assess the quality of the scoring functions, we also propose a novel weighted extension to established rank correlation~\citep{Goodman:1954:MAC}, based on agreement with respect to the structured ground truth ranking.

A second key contribution is explicit quantification of the subjectivity or ``softness'' of soft attributes.
We identify ways to differentiate more from less subjective soft attributes, and for measuring how this affects item scoring. This result also has ramifications for standard tagging, drawing attention to the role of subjectivity.

As our third main contribution, we
address the problem of critiquing based on soft attributes, and
present empirical evidence to demonstrate the importance of debiased collection of ground truth. 
In particular, \emph{soft attribute-based critiquing} is the task of ranking items \emph{relative} to a given anchor item with respect to a given soft attribute.
Three families of methods are devised for this task (unsupervised, weakly supervised, and fully supervised), which correspond to increasingly advanced approaches. 
These methods are compared on two test collections: 
one created based on existing social tags and another constructed using our proposed approach.
We observe a discrepancy between the results on the two test collections, showing that the tag-based test collection is blind to item ranking improvements, making progress in this area difficult. This demonstrates how new datasets and evaluation approaches may enable significant progress in nuanced interpretation of natural language critiques, which could not previously be reliably addressed.
We also analyze performance with respect to attribute ``softness'' and find that methods perform significantly better on attributes with higher agreement as opposed to those with low agreement.

In summary, the main novelty of the present work is the formalization of the notion of soft attributes, which opens up new possibilities for more natural interactions with conversational recommender systems. 
Our main technical contributions are threefold:
(1) We present an efficient method for debiased collection of ground truth for comparing items with respect to a given soft attribute and release a new dataset (Section~\ref{sec:testcoll});
(2) We introduce a measure to quantify soft attribute subjectivity, and perform an analysis of ``softness'' (Section~\ref{sec:analysis});
(3) We introduce and formalize a much refined task of critiquing based on soft attributes, developing unsupervised, weakly supervised, and fully supervised approaches (Sections~\ref{sec:methods}--\ref{sec:results}).

%% file: 02-related.tex
\vspace*{-1.5\baselineskip}
\section{Related work}
\vspace*{-0.25\baselineskip}

We present an overview of key past work on conversational recommendation, then specifically on critiquing. Following this, we present connections to opinion mining and related item tagging.\\

\vspace*{-0.75\baselineskip}
\noindent
\textbf{\emph{Conversational Recommender Systems.}}
Conversation has become recognized as a key modality for recommender systems (RSs) \cite{WuSanner19,Radlinski:2019:CCP,Lei:2020:Estimation}. Under the broader umbrella of information seeking and recommendation, conversational interaction has been recognized as of particular interest to the research community \citep{Culpepper:2018:RFI}.
Growing out of early work on slot filling \cite{Williams:2016:DST} and facted search \cite{Tunkelang:2009:Faceted}, conversational recommendation is distinct in that the sequence of exchanges between the user and the system is less rigid in structure, and often allow natural language dialogue. \citet{Radlinski:2017:TFC} postulated specific desirable properties of conversational search and recommendation systems in general, with critiquing being a core property. 
Various aspects of conversation have been addressed---including selecting preference elicitation questions to ask \cite{Christakopoulou:2016:TCR,Sepliarskaia:2018:Elicitation}, deep reinforcement learning models to understand user responses \cite{Sun:2018:CRS}, multi-memory neural architectures to model preferences over attributes \cite{Zhang:2018:TCS}, and neural models for recommendation directly based on conversations \cite{Li:2018:TDC}. Our work continues in this thread, albeit with a focus on semantic understanding of user utterances, at a level of detail that has not been addressed before. \\

\vspace*{-0.25\baselineskip}
\noindent
\emph{\textbf{Critiquing in Recommender Systems.}}
A specific interaction in conversational recommendation, critiquing seeks user reactions to items or sets of items. As such, critiquing-based RSs make recommendations, then elicit feedback in the form of critiques~\citep{Chen:2012:CRS}.  
The user may give feedback on various facets of importance, e.g.~the airline and cost of a flight, or time and date of travel, with respect to options presented~\citep{Linden:1997:IAU}. 
This is often repeated multiple times before the user makes a final selection.
\citet{Chen:2006:ECR} present a user interface facilitating critiquing, by letting users indicate how individual attributes should be changed (e.g., different screen size for a digital camera).
Critiquing has also been studied in conversational RSs~\citep{McCarthy:2010:ECR,WuSanner19} whereby users can affect recommendations along some standard item attributes with which items have been labeled, e.g.,~genres in movies.
\citet{Bi:2019:CPS} present a product search model that incorporates negative feedback on certain item properties (aspect-value pairs).  Unlike them, we solicit unconstrained natural language feedback, not limited to specific item properties.
Critiquing also happens in human conversations, where it is also not limited to predefined facets~\citep{Radlinski:2017:TFC}.  Being able to interpret soft attributes, which is our main objective, would thus enable the development of conversational agents that more closely resemble a human-to-human conversation.
Some previous work has also modelled how to learn different users' different definitions for particular terms. For instance, a \emph{safe} car may mean different things to different people~\cite{Boutilier:2009:Online,Boutilier:2010:Simultaneous}, although evaluation was limited to simulations. 

A recent line of work aims to facilitate language-based critiquing in modern embedding-based recommender systems beyond explicitly known item attributes~\citep{Wu:2019:DLB,Luo:2020:LLC,Luo:2020:DCV,Li:2020:ROA}.  The central theme of these efforts is to co-embed subjective item descriptions (i.e., keyphrases from user reviews) with general user preference information.  Then, end-to-end recommender architectures are trained to suppress items that match a critiqued keyphrase.  We also address the problem of natural language critiquing, but our focus lies on measuring the \emph{degree} to which a soft attribute applies to a given item, as opposed to end-to-end recommender architectures.  Our ranking models also operate in a latent embedding space to help overcome data sparsity and to account for variety in language usage, but we represent soft attributes as a function over item embedding dimensions, instead of trying to co-embed them in the same space. This way we are not constrained by keyphrase extraction.

\noindent
\emph{\textbf{Comparative Opinion Mining.}} 
Not to be confused with opinion mining, \emph{comparative opinion mining} deals with identifying and extracting information that is expressed in a comparative form~\citep{Varathan:2017:COM}.
Long studied by linguists, \citet{Jindal:2006:MCS} presented the first computational approach to comparative sentence extraction. 
Once comparative sentences are identified, comparative elements must be extracted: the entities and the attribute/aspect on which they are being compared, the comparative predicate, and the comparison polarity.  
For instance, \citet{Jindal:2006:MCS} identify entities by POS tagging, and
recognize comparative predicates using a manually compiled list.  
More recently, \citet{Kessler:2013:DPC} employ semantic role labeling techniques instead. 
Comparative sentences may be utilized in various ways, e.g.~for determining which of two entities is better overall~\citep{Kurashima:2008:REU,Zhang:2013:PCN,Tkachenko:2014:GME} or obtaining a global ranking of entities on a given aspect~\citep{Li:2011:PCU}.
A typical approach is to build a directed graph of entities, where edge weights encode the degree of belief that one entity is better than the other on a given aspect.
Then, entities can be ranked by some measure of graph centrality~\citep{Li:2011:PCU,Kurashima:2008:REU}.

However, the aspects considered previously are limited, often coming from a fixed ontology (e.g.,~ \cite{Li:2011:PCU}). 
Further, comparative opinion statements in natural text are uncommon, with estimates that 10\% of sentences in typical reviews contain a comparison~\cite{Kessler:2014:CCP}.
The most important difference in our work is that we aim to interpret arbitrary critiques, allowing direct navigation of the recommendation space. By designing a data collection and evaluation specifically for this task, we do not limit ourselves to common terms nor to reviews, but rather to natural critiques.

%% file: 03-testcoll.tex
\section{Building a Test Collection for Critiquing based on Soft Attributes}
\label{sec:testcoll}

We begin this section by motivating and formally introducing the notion of \emph{soft attributes} (Sec.~\ref{sec:testcoll:problem}).
Next, we describe the process of creating a reusable test collection for determining how a soft attribute applies to given items.  At a high level, the process consists of (1) sampling soft attributes, (2) obtaining ground truth item orderings for each attributes, and (3) developing an appropriate evaluation measure for this task.
A main challenge is that obtaining complete and universal ground truth orderings of items is infeasible, because of the human annotation effort required, and because people often disagree on the relative ordering of items.
We therefore use two types of evaluation collections. The first, which we term the MovieLens Attribute Collection, reuses past social tag annotations, with all their pertinent limitations and biases (Section~\ref{sec:eval:test_silver}). Our second dataset, which we term the Soft Attribute Collection, is collected with a process designed to reduce the biases (Section~\ref{sec:eval:test_gold}).

\subsection{Defining Soft Attributes}
\label{sec:testcoll:problem}

Consider the partial exchange between a user and an agent shown in Table \ref{tab:sample}.
The terminology employed and the aspects of this particular movie that attract the user's interest are likely to be useful for an RS providing future recommendations. Here, the user mentions concepts such as the \emph{immersiveness} of the movie, the \emph{in-depth exploration of a particular time period}, a \emph{bumbly character} that they relate to, and the \emph{level of violence} depicted. Yet it is difficult to argue that many of these attributes could be attached as definitive labels (i.e., tags) for the movie. Rather, the concepts are \emph{soft} attributes.

\begin{table}
	\footnotesize
	\caption{Example partial exchange between a user and an agent about the user's movies interests
	\cite{Radlinski:2019:CCP}.}
	\label{tab:sample}
	\vspace*{-\baselineskip}
	\begin{tabular}{lp{7cm}}
	\tt USER	    &\tt Zodiac's one of my favorite movies.\\
	\tt USER 	    &\tt Zodiac the movie about a serial killer from the '60s or '70s, around there.\\
	\tt ASSISTANT	&\tt Zodiac? Oh wow ok, what do you like about that movie?\\
	\tt USER	    &\tt And I just think serial killers, in general, are interesting, so the movie was really good. And it was just neat to see that world. Like it went really in-depth. It was like almost 3 hours long, so you got to really feel like you were a part of that world and time period. And see what the detectives and the police did and the investigators.\\
	\tt ASSISTANT	&\tt So you feel like you were part of that world ?\\
	\tt USER	    &\tt Yeah. It was really an immersive movie.\\
	\tt ASSISTANT	&\tt If you were in the movie what character do you feel most relatable\\
	\tt USER	    &\tt Probably the main character, Robert Graysmith, just cuz he's awkward and bumbly. So, I guess that's the one I would be the closest to.\\
	\tt ASSISTANT	&\tt What scene do you like the best?\\
	\tt USER	    &\tt Probably the most memorable one is the murder at the lake. Just cuz it's really vivid and horrific to watch. But it's very memorable.\\
	\end{tabular}
	\vspace*{-\baselineskip}
\end{table}

\begin{definition*}
We define a \textbf{soft attribute} as a property of an item that is not a verifiable fact that can be universally agreed upon, and where it is meaningful to compare two items and say that one item has more of the attribute than another.
\end{definition*}
\noindent
A number of examples of soft attributes are provided in Table~\ref{tbl:softattr} (Soft Attribute Collection).
We highlight two main distinguishing features that make a given attribute \emph{soft}.
First, there is commonly a question of degree. For instance, consider the movie attribute ``violent.'' If taken to either apply or not apply to a particular movie, it would be natural to ask what it takes for a movie to be marked as such. Are there different types of violence? How much violence is necessary in a movie before an entire movie is considered violent? Thus it is critical to model the degree to which each soft attribute applies to a given item.
Second, people may disagree in their assessment, even with a real-valued measure. While there are societal norms for attributes like violence (reflected in movie rating criteria, corresponding to a rating scale), this need not apply in general. In fact, we may ask whose opinion about an attribute (such as ``violence'') should matter if a given user says that they prefer ``less violent movies''? As such, different people may have different norms, expectations and thresholds for a given soft attribute.  
\citet{Boutilier:2009:Online,Boutilier:2010:Simultaneous} for instance postulated that an attribute such as ``safety'' may have different sub-aspects, with different people placing different importance on these sub-aspects.
We also include soft attributes that do not have norms: for example, ``exciting'' and ``boring'' are attributes where different people may entirely disagree. 

Another important difference of soft attributes versus social tags is that most tagging approaches intentionally bias users towards a consistent vocabulary, for example displaying frequent past labels \cite{Vig:2010:MovieLensTags,Harper:2015:MDH} or pre-defining an ontology of genres \cite{GoodReads:2018}. This encourages social taggers to conflate different semantics in popular tags.

We also note that soft attributes do not need to necessarily apply to all items in the collection. For example, ``realistic CGI'' in a movie is a soft attribute, yet CGI is not always present. This means that for any given soft attributes there is a (personal) partial order over items, where some items have the attribute more or less than others, while others are incomparable.

\subsection{Items Considered}
\label{sec:testcoll:items}

Here, we work within the movies domain as it has received the most attention in recommender systems research, and has a rich variety of publicly available resources~(e.g.,~\citep{Harper:2015:MDH,Radlinski:2019:CCP}). 
However, we note that there is nothing domain dependent in our approach and the same procedure may be employed to create test collections for other domains. 
Our corpus consists of the 300 most popular movies in the MovieLens-20M collection \citep{Harper:2015:MDH}, where popularity is measured by the number of users who rated the movie. We annotate each movie with its reviews in the Amazon review corpus~\cite{AmazonReviews:2015}, using the mappings established in \citep{Zemlyanskiy:2021:DLS}.\!\footnote{\url{http://goo.gle/research-docent}}

\begin{table}
\small
\centering
    \caption{Examples of soft attributes from our test collections.}
    \label{tbl:softattr}
    \vspace*{-\baselineskip}
    \begin{tabular}{cccc}
    \toprule
    \multicolumn{4}{c}{\bf MovieLens Attribute Collection} \\
    \midrule 
        action & comedy & depressing & dystopia \\
        romance & sci-fi & superhero & suspense \\
    \midrule 
    \multicolumn{4}{c}{\bf Soft Attribute Collection} \\
    \midrule 
        artsy & feels real & incomprehensible & intense \\ 
        light-hearted & predictable & simplistic script & violent \\ 
    \bottomrule
\end{tabular}
\vspace*{-0.5\baselineskip}
\end{table}

\subsection{MovieLens Attribute Collection}
\label{sec:eval:test_silver}

As frequently done~\citep{Sigurbjornsson:2007:FTR,Carman:2008:TDP,Rendle:2010:PIT}, one can create a synthetic test collection based on social tags (such as those present in MovieLens).  We do so, noting that its main limitation is that it does not allow for to measurement of the  ``degree'' of a soft attribute, only its presence or absence.  We shall investigate later on (in Section~\ref{sec:results}) how this affects the observed performance of ranking methods.

\subsubsection{Soft Attributes}

The top 100 most frequent MovieLens tags are taken as soft attributes, excluding tags that are named entities (names of actors, directors, studios, etc.), refer to adult content, or contain coarse language. We note that each of the tags has been assigned to 200--900 different items.  Table~\ref{tbl:softattr} shows examples.

\subsubsection{Ground Truth}

We identify two sets of items: those that have not been assigned that tag by any user, $\mathcal{X}^-$, and those where a significant portion $\alpha$ of users who assigned any tag to an item assigned the given tag, $\mathcal{X}^+$.
We take the threshold to be $\alpha = 0.15$. This is a conservative threshold, leaving typically only a handful of items as positive examples for each tag. Arguably, more items could be included, but that would come at the expense of potentially introducing noise.  A manual inspection of the results showed that all items in $\mathcal{X}^+$ appear to be valid examples for the tag.  Because of this conservative threshold, there were no positive examples for 18 tags, leaving us with 82 tags as soft attributes.

For reliable statistics, we also require items to be tagged by at least 50 users. This leaves 238 out of the original 300 movies in this test collection.
Table~\ref{tab:silver_stat} presents descriptive statistics for the dataset.

\subsubsection{Evaluation Measure}
To be able to evaluate algorithms that score items with respect to soft attributes, we need a measure that defines the goodness of a ranking.
Our evaluation measure is based on the number of concordant and discordant pairs between the generated ranking and the ground truth, specifically Goodman and Kruskal's gamma~\citep{Goodman:1954:MAC} as a measure of rank correlation:
\begin{equation}
    G = \frac{N_s-N_d}{N_s+N_d} ~, \label{eq:g}
\end{equation}
where $N_s$ is the number of concordant pairs and $N_d$ is the number of discordant pairs with respect to the ground truth ordering. $G$ can range from $-1$ (perfect inversion) to $+1$ (perfect agreement).

\begin{table}
\small
\centering
  \caption{Descriptive statistics of the test collections.}
  \label{tab:silver_stat}
  \vspace*{-\baselineskip}
  \begin{tabular}{lr}
    \toprule
    \multicolumn{2}{c}{\bf MovieLens Attribute Collection} \\
    \cmidrule{1-2} 
    Description & \!\!\!\!Count (min/mean/max) \\
    \cmidrule{1-2} 
    Number of soft attributes & 82 \\
    Positive examples / attribute & 1 / 5.77 / 28 \\
    Negative examples / attribute & 144 / 214.7 / 234 \\
    \midrule 

    \multicolumn{2}{c}{\bf Soft Attribute Collection}
    \\
    \cmidrule{1-2}
    Description & Count\\
    \cmidrule{1-2}
    Number of soft attributes & 60\\
    Sets of movies rated & 5,991 \\
    Pairwise preferences & 249,863 with 52,352 ties\\
    \bottomrule
\end{tabular}
\end{table}

\subsection{Soft Attribute Collection}
\label{sec:eval:test_gold}

While the MovieLens Attribute Collection was created using a common approach, tags suffer from a number of biases. Not least, item popularity biases the potential for items to receive particular tags, and the very nature of tagging limits the set of assigned tags to a handful of most strongly associated ones.  Yet, the fact that a tag has not been assigned to an item does not necessarily mean that it does not apply to that item at all.
Here, we present a multi-step approach that aims to reduce such biases. The final dataset statistics are presented in Table \ref{tab:silver_stat}, with the data available for download.\!\footnote{\url{http://goo.gle/soft-attribute-data}}

\subsubsection{Soft Attributes}

We sample soft attributes from the CCPE-M dataset~\citep{Radlinski:2019:CCP}. It 
consists of over 500 English dialogs between a user and an assistant discussing movie preferences in natural language.  Utterances are annotated with (1) entity mentions, (2) preferences about entities, (3) descriptions of entities, and (4) other statements of entities.  Of these, we leverage annotations of preferences (2).  First, we filter for utterances that potentially express a preference, by simply checking for the presence of the words ``more,'' ``less,'' or ``too.''  The resulting utterances were evaluated manually, and soft attributes identified.  A conservative filtering was applied, removing expressions that were problematic even for humans to interpret.  This yielded 173 unique soft attributes, of which we sample 60 for our evaluation (the probability of selecting an attribute was proportional to its frequency of use).  Table~\ref{tbl:softattr} lists a few examples.

\subsubsection{Ground Truth}
\label{sec:ground_truth}

Our methodology for collecting ground truth for soft attributes is one of the key contributions of this work. In particular, we focus on efficient collection of relative preferences over pairs of items for the same soft attribute, to enable inter-rater effects to be fairly measured. Additionally, the rater interface is designed to reduce popularity biases, as well as biases due to certain soft attributes never even being considered for certain items. \\

\noindent
\emph{\textbf{Annotation Procedure.}}
\label{sec:eval:annot}
For these attributes, we design a two-stage procedure to obtain pairwise item orderings via crowdsourcing. In particular, we recruited native English speakers based in the United States and Canada with a record of high quality annotation. 

In Stage 1, the crowd workers are asked to indicate which movies they have seen, from a pool of popular movies (cf. Section~\ref{sec:testcoll:items}). 

\begin{figure}[t]
    \centering
    \includegraphics[width=7cm]{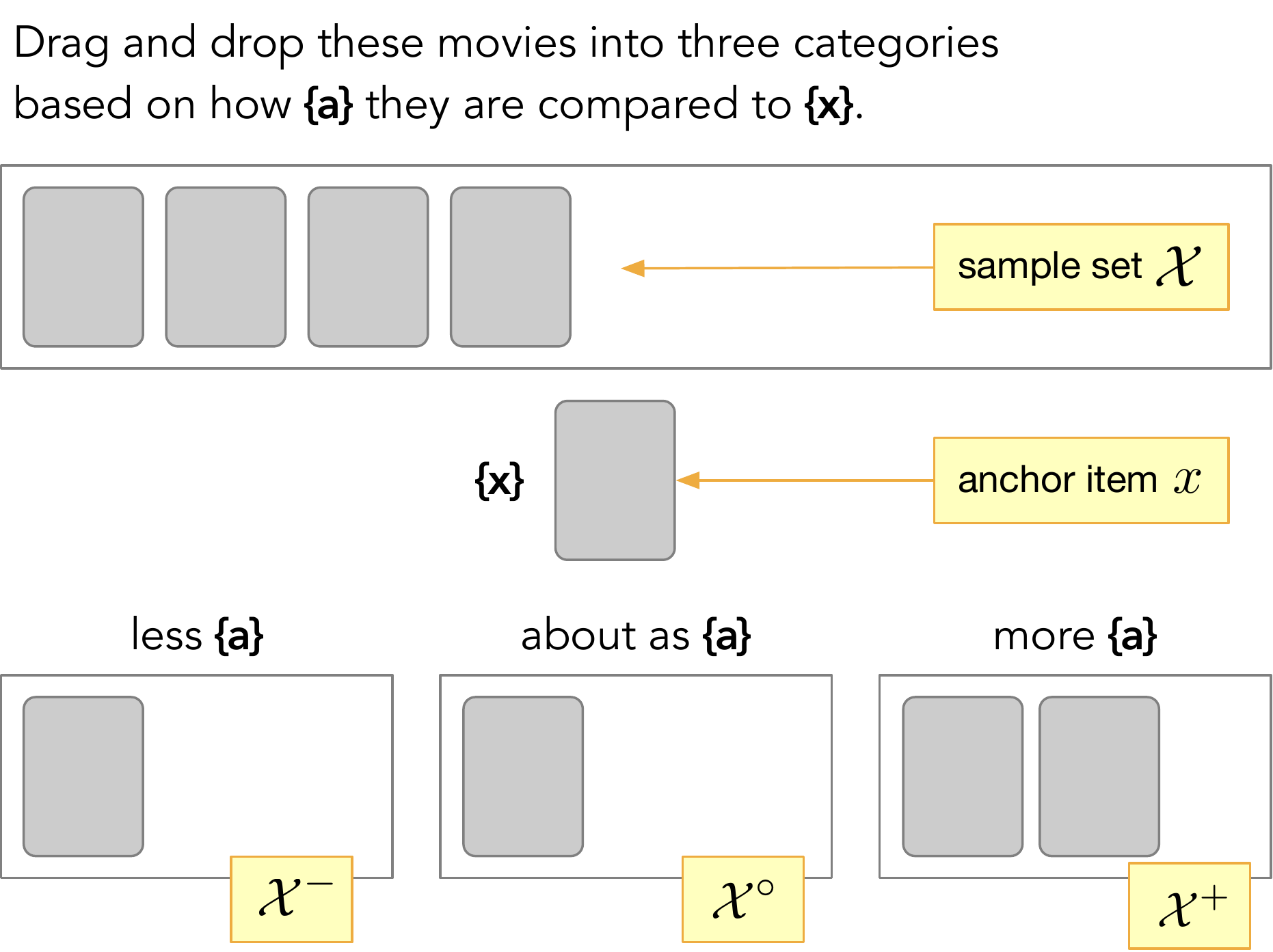}
    \vspace*{-0.8\baselineskip}
    \caption{User interface for obtaining item orderings. Given a soft attribute ($a$) and a set of items ($\mathcal{X}$), workers are required to drag the image for each item based on how this item compares to the anchor item ($x$) with respect to $a$. Items are thus partitioned into three classes, $\mathcal{X^-}, \mathcal{X^\circ}$ and $\mathcal{X^+}.$}
    \label{fig:ui}
    \vspace*{-1\baselineskip}
\end{figure}

In Stage 2, each worker is presented with a specific ordering task for each soft attribute $a$. As illustrated in Figure \ref{fig:ui}, they are provided with an anchor item $x$ and a small sample set of items (usually 10), which are to be divided into three categories with reference to the anchor item: ``less $a$ than $x$,'' ``about the same $a$ as $x$,'' and ``more $a$ than $x$.''  We represent each item by its image (movie poster) and title. 
The set of items shown (both the anchor item and the sample set) are personalized for each worker, that is, they are chosen from the set of movies that the person has indicated as seen in Stage 1 so as to ensure that all judgments compare items on an equal footing.
By forcing workers to consider consistent terminology across a variety of items, we reduce biases due to the context in which individual items are usually considered, and typical comparisons which are made (e.g., dominated by the genres to which items belong, or tags given by other users in the past). 
By offering the ``about the same'' label, we also force an explicit decision as to whether a difference between two items is meaningful or not. This full set of judgments then yields pairwise preferences over all pairs of items placed into two different classes, as well as relative to the anchor item. \\

\vspace*{-0.5\baselineskip}
\noindent
\emph{\textbf{Sampling Methodology.}}
\label{sec:eval:sampling}
Notice that the way items are selected for crowd workers influences the number of pairwise item preferences that can be obtained.  To maximize this value of the rating task, we develop the following sampling methodology.
To increase the likelihood of meaningful comparisons between items, we score items using two baseline algorithms: the term-based item-centric and review-centric models (described in Section \ref{sec:term-based-ranking}).  First, we eliminate all items that the user has not marked as seen in Stage 1.  All seen items are then sorted by score and partitioned into $M\! =\! 5$ bins for each baseline, each with about the same number of items.

For a particular user and attribute $a$, we must select the anchor item $x$ and subset $\mathcal{X}$ to rank. 
The choice of anchor item $x$ is particularly key: if it is at an extreme for the soft attribute, most items that $x$ is compared to would fall into just one or two buckets, yielding few pairwise preferences. Therefore, only items not in the first or last bin for either baseline shall be selected as $x$. Second, as the anchor item is present in the most pairwise preferences inferred, we wish to bias it towards more popular items, maximizing the number of observed preferences for this attribute for different workers. Thus, from the set of possible anchor items, we sample the anchor item with probability proportional to the number of crowd workers who indicated that they had seen this item in Stage~1.
Finally, $\mathcal{X}$ is a stratified sample of one item from each of the $M$ bins for each baseline algorithm, sampling items with probability proportional to the number of workers who marked each item as seen in Stage~1. This gives $|\mathcal{X}|\! =\! 10$ items, unless the user has seen too few items.

\subsubsection{Evaluation Measure}

As explained above and in Figure~\ref{fig:ui}, raters (crowd workers) are presented, for each soft attribute, with a reference item $x$ and a sample set of items $\mathcal{X}$, to be categorized into three disjoint sets: $\mathcal{X}^-$ (``less $a$ than $x$''), $\mathcal{X}^\circ$ (``about the same $a$ as $x$''), and $\mathcal{X}^+$ (``more $a$ than $x$'').
A standard approach for evaluating scoring algorithms given pairwise preferences is to count how many pairs are ordered in agreement with the test data. 

However, with three classes, we are able to measure agreement with a stronger metric. We present an extended version of the pairwise agreement metric that differentiates between more/less and \emph{much} more/less.
Specifically, we define a weighted extension of Goodman and Kruskal's gamma rank correlation measure \cite{Goodman:1954:MAC}:
\begingroup
\setlength\belowdisplayskip{1pt}
\setlength\belowdisplayskip{2pt}
\begin{equation}
    G' = \frac{(N_s-N_d) + 2(N_{ss}-N_{dd})}{(N_s+N_d) + 2(N_{ss}+N_{dd})} ~, \label{eq:gprime}
\end{equation}
\endgroup
where the number of concordant and discordant pairs, $N_s$ and $N_d$, are measured on elements in adjacent buckets, that is, $\mathcal{X}^-$ vs. $\mathcal{X}^0 \cup \{ x \}$ and in $\mathcal{X}^0 \cup \{ x \}$ vs. $\mathcal{X}^+$ (the anchor item $x$ is considered in the ``middle'' bucket).\footnote{Measurement on elements in buckets $\mathcal{X}^a$ vs. $\mathcal{X}^b$ means that all pairs of items $x_a \in \mathcal{X}^a, x_b \in \mathcal{X}^b $ are considered; if $x_a$ is ranked higher (i.e., given a higher score) than $x_b$ in the ranking, then it counts as a concordant pair, otherwise as a discordant pair.}

For pairs of items judged \emph{much} more/less (in buckets $\mathcal{X}^-$ vs. $\mathcal{X}^+$), we give double weight:
$N_{ss}$ counts the number of pairs of items in $\mathcal{X}^-$ vs. $\mathcal{X}^+$ that are ranked identically, while $N_{dd}$ counts how many of those are ranked differently, in a given item ranking as compared to the ground truth.
Similar to the original $G$ (Eq.~\eqref{eq:g}), the extended $G'$ also ranges from -1 (perfect anti-correlation) to +1 (perfect correlation), with a score of 0 indicating random agreement.

In the rare case where all items are rated into a single bucket, namely to ``about the same $a$ as $x$,'' this metric is undefined. We leave out these ratings from the evaluation.

Note that unlike the original $G$, our extended $G'$ does not assume an universally accepted ground truth. Instead, it measures how well a ranking aligns with the preferences of a single rater. In our experiments, we report on mean $G'$ values over all raters.

%% file: 04-analysis.tex
\vspace*{-0.25\baselineskip}
\section{Analysis of Soft Attributes}
\label{sec:analysis}
\vspace*{-0.25\baselineskip}

There are two particularly fundamental concepts when considering a given soft attribute: (1) How subjective is the meaning of the \emph{text} used for that attribute when applied to particular items?, and (2) Do we observe significant ``equality?'' That is, when a user requests an item with a given attribute changed, how much change is expected? We address what our data tells us about each in turn.

\subsection{Quantifying Subjectivity}
\label{sec:analysis:consistency}

Past attribute datasets have not analyzed personal variations in the \emph{meaning} of a term, or in the \emph{relative applicability} of soft attributes: While often scored for relevance (as in \cite{Vig:2010:MovieLensTags}), it is difficult to argue that scores are an unbiased measure of disagreement about meaning. In our work, these questions are reflected in inter-judge agreement: Do different people considering the same soft attribute for the same items agree on which is ``more''?  

For example, consider the attribute ``predictable.'' 
Different people may consider different plot devices predictable, based on their past experience, leading to  disagreement about the relative ordering of some movies. On the other hand, we might consider the attribute ``funny'' under-specified: for some people it may indicate slapstick comedy, while for others it may indicate peculiar characters, plots, and settings (often characteristic of British comedy). As such, disagreements may come from different definitions. 

To measure soft attribute subjectivity, we identify all pairs of movies $(x, x')$ that have been ranked for the same attribute~$a$ by different raters. At this point, recall that multiple judgment of tuples $(a, x, x')$ is not something we can directly control.  Rather, raters specify which movies they have seen, and anchor items and movies are sampled as described above---meaning we rely on this being an unbiased sample of such tuples.  Most tuples were judged just twice.

We define preference agreement by considering every pair of judgments for every soft attribute. Agreement is defined as the fraction of pairs where either (1) both raters agree on the direction of preference, or (2) at least one rater indicates a lack of preference (i.e.,~there is no disagreement).  Formally, let $N^a_{x \prec x'}$ be the number of raters who indicated that attribute $a$ applies more to $x'$ than to $x$, and $N^a_{x\approx x'}$ that for raters who indicated that $a$ applies about equally. For each pair of movies $x$ and $x'$,
\begin{equation}
\begin{array}{cc}
	p^a_{x \prec x'} = \frac{N^a_{x \prec x'}}{N^a_{x\prec x'} + N^a_{x=x'} + N^a_{x \succ x'}} 
	\quad
	p^a_{x\approx x'} = \frac{N^a_{x\approx x'}}{N^a_{x\prec x'} + N^a_{x=x'} + N^a_{x \succ x'}}
\\ 
	p^a_{x\succ x'} = \frac{N^a_{x\succ x'}}{N^a_{x\prec x'} + N^a_{x=x'} + N^a_{x \succ x'}} 
\end{array}
\end{equation}
For attribute $a$, $\mathit{agree}(a) = \mathit{mean}_{x,x'}[\mathit{agree}(a,x,x')]$, 
where 
\begin{eqnarray}
	\mathit{agree}(a, x, x') & = & p^a_{x\prec x'} \cdot \left(p^a_{x\prec x'} + p^a_{x\approx x'}\right) \\
	 & & + p^a_{x\succ x'} \cdot \left(p^a_{x\succ x'} + p^a_{x\approx x'}\right) + p^a_{x\approx x'} , \nonumber
\end{eqnarray}
\noindent
taking lack of preference $(x\approx x')$ as indicating agreement with either direction. Based on agreement rate, we divide the soft attributes into three equal-sized groups: High, Medium, and Low agreement attributes.  A sample from each is shown in Table~\ref{tab:agreement}.
We see that the attributes with highest score are reminiscent of typical tags in tag corpora for movies (we found it surprising that \emph{funny} has high agreement). Many of the attributes with lowest agreement relate to personal preferences (\emph{entertaining}, \emph{overrated}). It is also particularly interesting to observe that attributes that are seemingly opposite (e.g., \emph{intense} and \emph{boring}) can have quite different agreement rates.

\begin{table}
    \caption{Selection of attributes with High, Medium, and Low agreement rates ($\mathit{agree}(a)$).}
    \label{tab:agreement}
    \vspace*{-0.75\baselineskip}
	\footnotesize
    \centering
    \begin{tabular}{ll@{}c@{~~~}c@{}r@{}r}
    \toprule
    Group & Attribute & $\mathit{agree}(a)$ & Distinct movie pairs & \multicolumn{2}{c}{Total comparisons}\\
    \midrule
High 
    & scary & 0.962 & 291 & \hspace{5mm}617 & (6 ties)\\
agreement 
    & gory & 0.952 & 246 & 513 & (3 ties)\\
    & action filled & 0.950 & 277 & 583 & (23 ties)\\
    & funny & 0.949 & 318 & 672 & (10 ties)\\
    & terrifying & 0.947 & 289 & 623 & (19 ties)\\
    & violent & 0.946 & 290 & 615 & (13 ties)\\
    & intense & 0.937 & 297 & 634 & (27 ties)\\
\midrule
Medium 
	& fictionalized & 0.894 & 243 & 511 & (12 ties)\\
agreement
    & tearful & 0.880 & 233 & 493 & (10 ties)\\
    & romantic & 0.885 & 272 & 585 & (21 ties)\\
    & confusing & 0.882 & 183 & 390 & (16 ties)\\
    & mushy mushy & 0.855 & 241 & 510 & (16 ties)\\
    & exaggerated & 0.830 & 288 & 606 & (24 ties)\\
\midrule
Low 
	& unique story & 0.813 & 225 & 470 & (22 ties)\\
agreement 
    & original & 0.808 & 189 & 391 & (20 ties)\\
    & entertaining & 0.796 & 270 & 569 & (18 ties)\\
    & boring & 0.791 & 234 & 514 & (28 ties)\\
    & dynamic & 0.785 & 280 & 590 & (25 ties)\\
    & overrated & 0.766 & 280 & 596 & (26 ties)\\
    \bottomrule
    \end{tabular}
    \vspace*{-0.75\baselineskip}
\end{table}

\subsection{Quantifying Equality}

As we tasked raters to bucket movies into three categories, we can observe the distribution over the counts in these buckets. Given an anchor item and attribute, on average raters put\footnote{These counts do not add up to 10 as $|\mathcal{X}|$ was not always 10.} 3.46 movies into $\mathcal{X}^-$, 3.28 movies into $\mathcal{X}^\circ$, and 3.19 movies into $\mathcal{X}^+$. While this distribution is likely influenced by the stratified sampling of items in $\mathcal{X}$, it confirms that there often exist many pairs of movies to which a given attribute applies equally---even when many others can be classified as more or less. Considering the scores of movies in the $\mathcal{X}^\circ$ set (``about the same $a$ as $x$''), it would be possible to identify thresholds that determine how different scores for items should be for a recommender to have satisfied a user's critique of more/less without necessitating extreme differences. For instance, asking for a ``less violent'' movie compared to a very violent one should likely not lead to recommendations of movies with no violence at all.

We also observe that \emph{long} (4.55 movies in $\mathcal{X}^\circ$), \emph{documentary style} (4.51 movies), \emph{well directed} (4.13 movies) and \emph{original} (4.13 movies) have the most items in $\mathcal{X}^\circ$. This suggests that critiques of these attributes are likely to eliminate many movies simply because significant differences are less common. In contrast, the attributes \emph{playful} (2.53 movies in $\mathcal{X}^\circ$), \emph{funny} (2.57 movies), \emph{sappy} (2.69 movies) and \emph{scary} (2.78 movies) are much more likely to have raters provide a more complete order over movies, thus fewer would be eliminated on the grounds of being too similar so as to satisfy a critique.

%% file: 05-methods.tex
\section{Scoring Items by Soft Attribute}
\label{sec:methods}

\begin{figure*}
\centering
\begin{minipage}{.5\textwidth}
  \centering
  \includegraphics[width=8cm,trim=0 10 0 0]{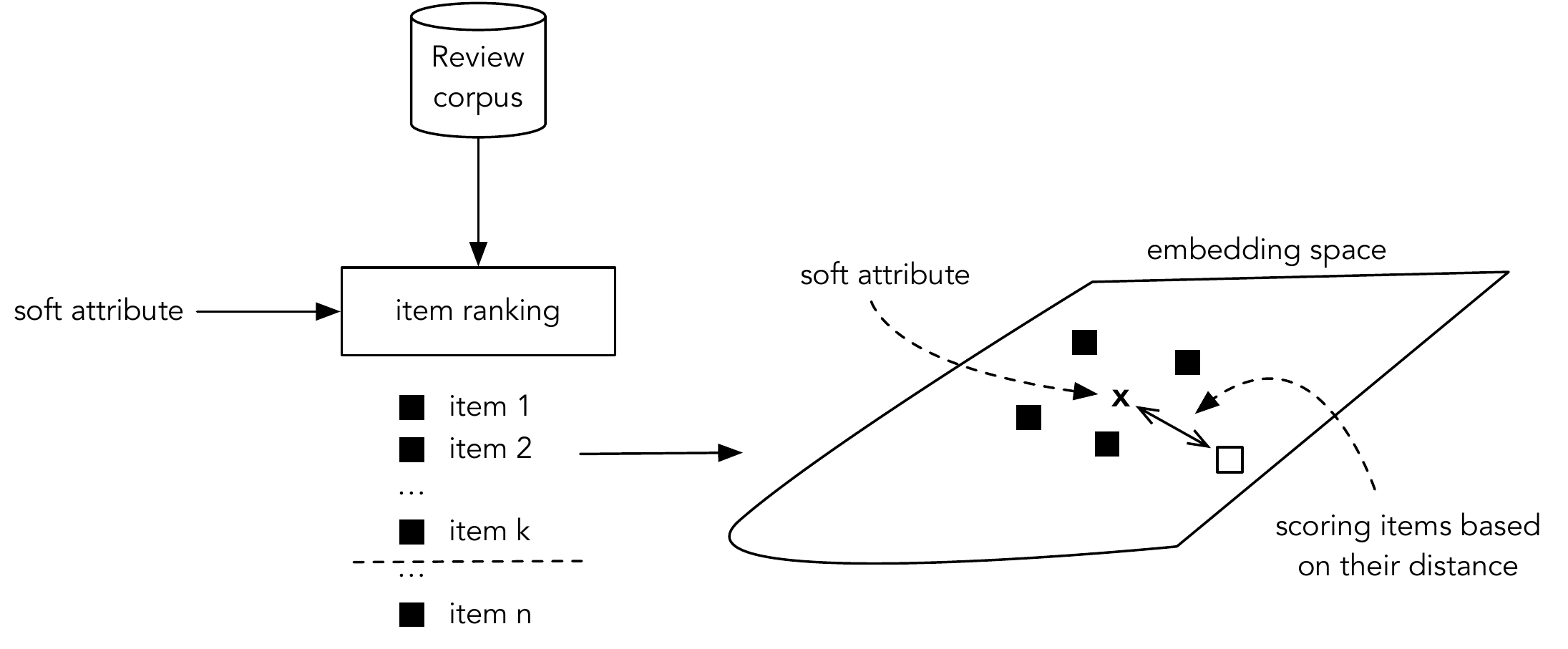}
  \caption{Unsupervised item ranking using the Centroid-\\Based (CB) method.}
  \label{fig:TB+CBE}
\end{minipage}%
\begin{minipage}{.5\textwidth}
  \centering
  \hspace{-1.5cm}
  \includegraphics[width=8cm,trim=0 10 0 0]{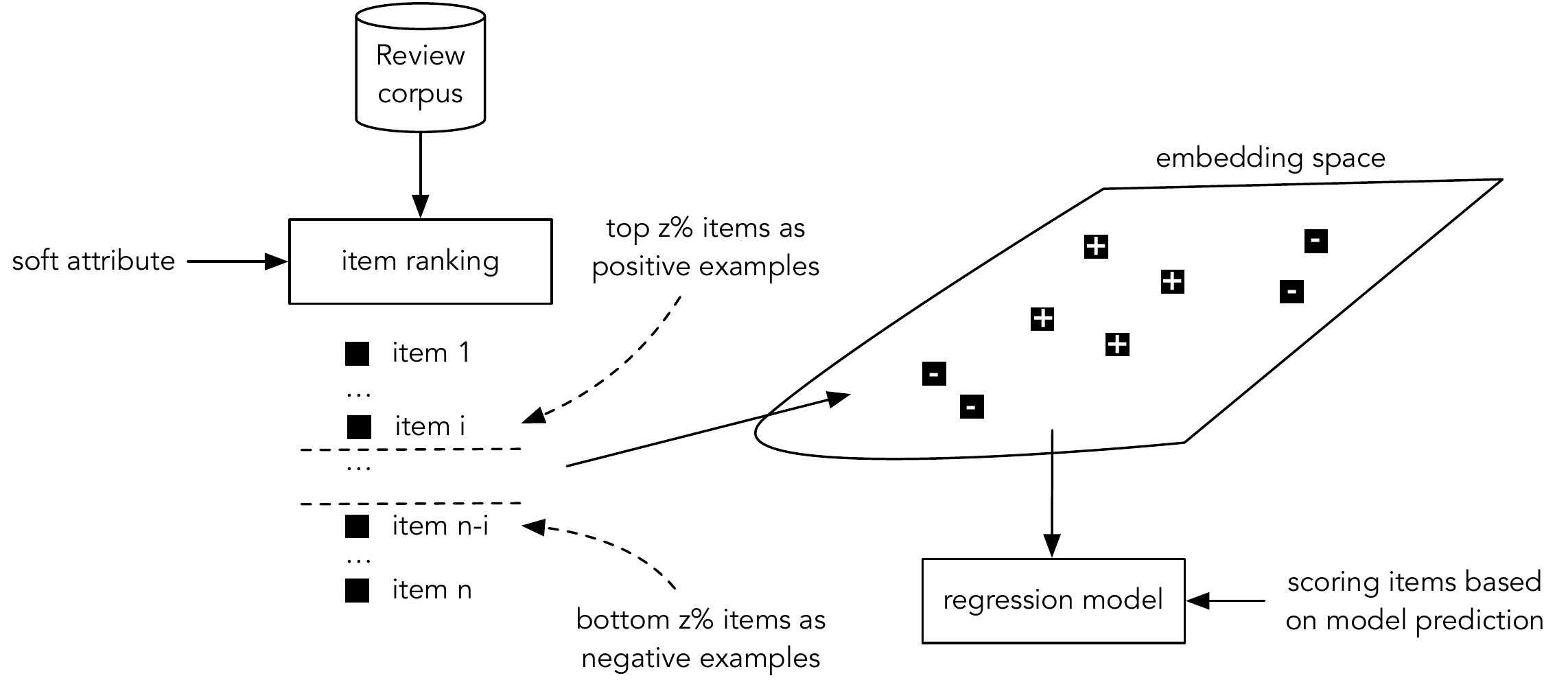}
  \caption{Weakly supervised item ranking using the Weakly-supervised Weighted Dimensions (WWD) method.}
  \label{fig:TB+WDE}
\end{minipage}
\vspace{-\baselineskip}
\end{figure*}

We now address the question of how to apply soft attribute critiques in a recommendation setting. Specifically, we cast this task as a ranking problem.  
Our approach hinges on the idea that for the type of critiquing we aim to support---specifying some soft attribute of the desired item(s) relative to a given anchor item---it is sufficient to determine the relative ordering of items. Once that relative ordering is established, we can locate the anchor item in the ranking and move in the required direction (and distance) with respect to that (e.g., ``little less'' or ``much more'').
Our formal objective then is to devise a scoring function $\mathit{score}(x, a)$, where $x$ is an item in the collection and $a$ is a soft attribute.  If an attribute cannot be applied at all to a given item, it should ideally be assigned a score of 0. 

At their core, the scoring functions boil down to the question of how items and soft attributes are represented.
We represent items in a latent space by learning embeddings from item-rating data (Sec.~\ref{sec:approach:embeddings}).
Our underlying assumption is that a representation of soft attributes within the same embedding space can be learned.  (Note that we do not assume interpretability of these embeddings.)
We learn to represent soft attributes in three fundamentally different ways, from \emph{unsupervised} to \emph{weakly supervised} to \emph{fully supervised}.
\begin{itemize}[leftmargin=*]
    \item When no explicit training data over soft attributes exists, we use implicit signals from item reviews, employing established models from entity retrieval.  Taking the centroid of top-ranked items, in the spirit of pseudo-relevance feedback, we representation soft attribute using an \emph{unsupervised} approach (Section~\ref{sec:approach:ranking_unsupervised}).
    \item Instead of considering only positive examples, we can also take the top-ranked and bottom-ranked items as positive and negative (pseudo-)training examples, respectively, to learn a pointwise regression model in a \emph{weakly supervised} way (Section~\ref{sec:approach:ranking_weak}).
    \item Finally, where pairwise item orderings for a given soft attribute are available (facilitated by the data collection methodology we introduce), we can use these pairwise preferences as training examples and learn a model using \emph{full supervision} (Section~\ref{sec:approach:ranking_full}).
\end{itemize}
We note that these models do not personalize the meaning of a soft attribute to a particular user, although we analyze the importance of personalization in Section~\ref{sec:results:attrs}.

\subsection{Generating Item Embeddings}
\label{sec:approach:embeddings}

Many methods have been proposed to compute item embeddings, e.g.,~\cite{BarkanK16}. One of the most common ways to compute item representations from collaborative filtering datasets is matrix factorization~\cite{Koren2009}. Matrix factorization for collaborative filtering works by using the user-item rating matrix $\mathbf{R}^{n\times m}$ for $n$ users and $m$ items,  factorizing this matrix to two low rank matrices containing the user embeddings $\mathbf{U}^{n\times d}$ and item embeddings $\mathbf{X}^{m\times d}$. To this end the following objective function is minimized: 
\begin{equation}
    (r_{ij} - \langle \mathbf{u}_i, \mathbf{x}_j \rangle)^2 + \lambda_1 \|\mathbf{u}_i\|^2 + \lambda_2 \|\mathbf{x}_j\|^2  ~,
\end{equation}
where $r_{ij}$ is the rating of item $j$ by user $i$, $\mathbf{u}_i$ is the $d$-dimensional embedding for user $i$ and $\mathbf{x}_j$ that for item $j$. Each user and item embedding vector corresponds to one row in the user and item embedding matrices.
The objective function is minimized using stochastic gradient descent. 

\subsection{Unsupervised Ranking}
\label{sec:approach:ranking_unsupervised}

We present two unsupervised ranking approaches as baselines: one operating in the term space (Section~\ref{sec:term-based-ranking}) and another operating in the embedding space (Section~\ref{sec:centroid-based-ranking}).

\subsubsection{Term-based Ranking}
\label{sec:term-based-ranking}

Adapting well-established evidence aggregation strategies for entity retrieval~\citep{Balog:2018:EOS,Zhang:2017:DPF}, we leverage the corpus of item reviews, using soft attributes as search queries, for our term-based (TB) methods. Items are represented by aggregating reviews following an \emph{item-centric} or a \emph{review-centric} strategy.

\begin{description}
	\item \textbf{Item-centric Method} A term-based representation $d_x$ is built for each item $x$ by concatenating all reviews mentioning $x$: $d_x = \bigcup_{r\in R_x} r$. The representations are then scored using standard text-based retrieval models (here: BM25):
		\begin{equation}
		    score^{\textit{item}}(x,a)=score_{BM25}(d_x,q) ~.    
		\end{equation}
	\item \textbf{Review-centric Method} Reviews $r \in \mathcal{R}$ are ranked using standard retrieval models (here: BM25). Then, for each item $x$, the retrieval scores of reviews that mention $x$ are aggregated:
		\begin{equation}
		    score^{\textit{review}}(x,a) = \sum_{r \in R_x} score_{BM25}(r,q) ~,
		\end{equation}
		where $R_x$ denotes the set of reviews that mention $x$. 
\end{description}

\subsubsection{Centroid-based Ranking}
\label{sec:centroid-based-ranking}

Our centroid-based (CB) ranking method considers the top ranked items, in the spirit of pseudo relevance feedback, as representative examples of the soft attribute. Then, the centroid of these top ranked items is taken as the representation of the soft attribute in the embedding space. Other items can then be scored by computing their distance to the centroid.

Let $\textbf{v}_a$ denote the representation of the soft attribute in the item embedding space (an N-dimensional vector).  
We obtain $\textbf{v}_a$ by first ranking items with respect to the soft attribute using a term-based model over review text, and then taking the centroid of the top-k ranked items' embeddings:
\begin{equation}
    \mathbf{v}_a[i] = \frac{1}{k}\sum_{x \in X_k} \mathbf{x}[i] ~, \label{eq:attr_vec_centroid}
\end{equation}
where $X_k$ denote the set of top-k ranked items and $\mathbf{x}[i]$ refers to the $i^\mathrm{th}$ dimension of embedding vector $\mathbf{x}$.
Once $\mathbf{v}_a$ has been computed, a given item $x$ in the collection is scored against it by computing the cosine similarity of their respective embedding vectors: 
$\mathit{score}^\textit{CB}(x,a) = \textit{cosine}(\mathbf{x}, \mathbf{a})$. See Fig.~\ref{fig:TB+CBE} for an illustration.

\subsection{Weakly Supervised Ranking}
\label{sec:approach:ranking_weak}

While the centroid-based method represents items in the embedding space, it considers all factors (dimensions) equally important.  Instead, we would like to \emph{learn} which factors in the embedding space encode a particular soft attribute.  In the absence of explicit training labels, we again use term-based models for obtaining an initial ranking of items. Then, the top- and bottom-ranked items are taken as positive and negative training examples, respectively, to learn a regression model. Items can then be scored by applying this model and taking the prediction probabilities as scores.

\vspace{2mm}
Specifically, we employ Logistic Regression (LR), with target labels $y_a$, where $y_a=1$ when attribute $a$ applies to item $x$, and $y_a=0$ otherwise. 
To obtain these labels for training the LR model, we operationalize the notion of an attribute applying to an item by considering the item's position in the ranking for that attribute in the term-based model (thereby essentially taking the term-based ranking as a weak supervision signal). Items in top $z$\% positions of the ranking are taken as positive examples (as long as their initial scores are greater than zero). The same number of items from the bottom are taken as negative. 
The embedding of the soft attribute can then be computed by fitting a LR of the item embeddings $\mathbf{x}$ to the item attribute labels $y_a$ (see Fig.~\ref{fig:TB+WDE} for an illustration): 
\begin{equation}
    p\left(y_a = 1|x\right) = \frac{1}{1+\exp{(-\mathbf{w}_a \cdot \mathbf{x}})} \label{eq:attr_logreg}
\end{equation}
Intuitively, the model parameters $\mathbf{w}_a$ that are computed by LR reflect the importance (weight) of each dimension in the item embeddings in predicting the soft attribute. In this way, we compute which dimensions (and to what degree) are encoding this attribute. 
This can be seen as distilling a representation of the soft attribute from the item embeddings. 
Naturally one can compute the scores of items $\mathbf{x}$ with respect to the attributes by using the corresponding LR model. Our \emph{Weakly-supervised Weighted Dimensions} (WWD) method thus ranks items according to:  
\begin{equation}
    \textit{score}^\textit{WWD}(x,a) = p\left(y_a = 1|x\right) 
\end{equation}

\subsection{Fully Supervised Ranking}
\label{sec:approach:ranking_full}

The weakly supervised method (WWD) learns weights over the embedding space dimensions using pseudo-relevance feedback based on the \emph{review} corpus. Here, we show how the explicit item orderings (i.e., data collected using the methodology we describe in Section~\ref{sec:eval:test_gold}) can be leveraged to efficiently learn this weighting directly from the \emph{ordinal labelling} produced by raters using a ranking algorithm trained on pairwise preferences.

Specifically, given ground truth judgments as described in Section \ref{sec:ground_truth}, we infer all possible pairwise preferences. For each judgment with reference item $r$ and labelled sets $\mathcal{X}^-$, $\mathcal{X}^\circ$, and $\mathcal{X}^+$ we infer the following sets of preferences between items $i$ and $j$:
\begin{enumerate}
\item $\{i \succ\!\!\succ j\},~\forall i \in \mathcal{X}^+, j\in \mathcal{X}^-$
\item $\{i\succ j\},~\forall i \in \mathcal{X}^+, j\in \mathcal{X}^\circ \bigcup \{r\} $
\item $\{i\succ j\},~\forall i \in \mathcal{X}^\circ \bigcup \{r\}, j\in \mathcal{X}^-$
\item $\{i\sim j\},~\forall i,j \in \mathcal{X}^\circ \bigcup \{r\}$
\end{enumerate}

\noindent
The fully supervised model (\emph{Supervised Weighed Dimensions}, SWD) then uses a linear ranking support vector machine\footnote{The hyperparameter $C$ was set to 1. We analyzed the effect of tuning hyperparameters and using non-linear models including gradient boosted decision trees, finding no meaningful improvements. We hypothesize this lack of sensitivity to the precise ranking model to be due to the limited amount of training data, leaving identifying the best performing model to future work as it is not key to this work.} \cite{joachims:2002:optimizing}. Briefly, each preference $p$ was transformed into a constraint: $i \succ j$ becomes $\textit{score}^{SWD}(x_i) \ge 1 + \textit{score}^{SWD}(x_j) - \xi_p$, and $i \succ\!\!\succ j$ becomes $\textit{score}^{SWD}(x_i) \ge 2 + \textit{score}^{SWD}(x_j) - \xi_p$, where $x_i$ denotes the embedding vector of item $i$. Then $\textit{score}^{SWD}(x_i) = w \cdot x_i$, where $w$ is learned by minimizing $\frac{1}{2} w\cdot w + C \sum_p \xi_p$.

As with the WWD model, this model represents a soft attribute by a direction in the embedding space ($w$), albeit trained using explicit pairwise preferences between items rather than based on terms extracted from reviews.

%% file: 06-results.tex
\section{Evaluation}
\label{sec:results}

We evaluate scoring algorithms based on how many pairs are ordered in agreement with the ground truth data.  Specifically, we use the original Goodman and Kruskal gamma ($G$, Eq.~\eqref{eq:g}) for the MovieLens Attribute Collection (MovieLens for short) and our modified version ($G'$, Eq.~\eqref{eq:gprime}) for the Soft Attributes Collection (SoftAttr for short).
Our focus is to verify whether the algorithmic improvements we would expect are indeed observed when they are evaluated against different test collections.

\subsection{Experimental Setup}

For the unsupervised (CB) and weakly-supervised (WWD) methods, we report on a single parameter setting, selecting that which performs best across the board.  Specifically, we use 25 dimensional embeddings; for the centroid-based methods we use top $k=5$ items; for weighted-dimensions we take $z=0.4$.\footnote{We also performed a sensitivity analysis of these parameters and observed stable performance across a broad range of values ($k\,\in\, [1..100]$, $z\,\in\, [0.1~..~0.4]$). Additionally, we experimented with the dimensionality of embeddings (between 25 and 100), finding that higher dimensionality somewhat reduces performance due to over-fitting effects that frequently manifest themselves in the outright performance of high-dimensional matrix factorization techniques fitted on (relatively) small data.} 

For the supervised method, SWD, we use 10-fold cross validation: We split  the raters into 10 groups, and evaluate each group by using only the remaining 9 groups to train the ranking model. We report average performance across the 10 folds.

\begin{table}[t]
  \caption{Soft attribute ranking results on the MovieLens Attribute Collection (MovieLens) and Soft Attributes Collection (SoftAttr). The best scores in each block are in bold.}
  \label{tab:results}
  \vspace*{-\baselineskip}
  \small  
  \begin{tabular}{l@{\hskip2pt}l@{\hskip0pt}c@{\hskip2pt}c}
    \toprule
    \multicolumn{2}{l}{\textbf{Method}} & \textbf{MovieLens} & \textbf{SoftAttr} \\
    & & (G) & (G') \\
    \midrule
    \multicolumn{4}{l}{\emph{Unsupervised methods}} \\
    \midrule
    (TB-IC) & Term-based, item-centric & \textbf{0.800} & 0.110 \\ 
    (TB-RC) & Term-based, review-centric & 0.733 & \textbf{0.136} \\ 
    (CB+TB-IC) & Centroid-based (w/ TB-IC) & 0.404 & 0.087 \\ 
    (CB+TB-RC) & Centroid-based (w/ TB-RC) & 0.471 & 0.101 \\ 
    \midrule
    \multicolumn{4}{l}{\emph{Weakly supervised methods}} \\
    \midrule
    (WWD+TB-IC) & Weakly-sup. w. dim. (w/ TB-IC) & \textbf{0.539} & 0.194 \\ 
    (WWD+TB-RC) &  Weakly-sup. w. dim. (w/ TB-RC) & 0.517 & \textbf{0.200} \\ 
    \midrule
    \multicolumn{4}{l}{\emph{Fully supervised method}} \\
    \midrule
    (SWD) & Supervised weighted dimensions & & \textbf{0.485} \\
    \bottomrule
\end{tabular}
\end{table}

\subsection{Results}

Our overall results are reported in Table~\ref{tab:results}, averaged over all soft attributes and all users. The best performing methods in the supervised and unsupervised categories are bolded for each collection.

First, observe the \emph{magnitude} of the rank correlation scores when evaluated on each collection. The simple term-based models perform remarkably well when assessed on MovieLens, to the extent that one could assume the problem of ranking items for a given soft attribute could be substantially solved with a straightforward model. Yet we postulate that this common formulation captured by the MovieLens collection, namely with performance determined by how well items \emph{with} a given tag can be distinguished from those \emph{without}, is misleading. On the SoftAttr collection we see much lower scores overall, indicating that this more accurate abstraction of the attribute ranking problem proves to be considerably harder.

Next we observe that the relative ordering of systems on the MovieLens collection is TB $>$ WWD+TB $>$ CBE+TB, while on SoftAttr collection it is WWD+TB $>$ TB $>$ CBE+TB, also emphasizing the importance of the task encoded in the data enabling meaningful progress.  Centroid-based ranking (CB+TB) is least performant in both cases.  However, the retrieval-based baseline (TB) performs much better on the MovieLens collection than the weakly supervised approach (WWD+TB). On the SoftAttr collection it is exactly the other way around. We hypothesize that one of the main reasons that WWD+TB performs much better than CB+TB is that WWD+TB considers positive as well as negative evidence (i.e.,~movies at the bottom of the ranking for a given attribute), while the CB+TB baseline only considers positive evidence (i.e.,~movies where the soft attribute can reliably be considered to apply).

Finally, we turn to the supervised method, Supervised Weighted Dimensions (SWD). We see that direct supervision based on relative preference judgments yields significant improvement in performance on the SoftAttr collection.

Together, these results show the key contribution that our new data collection methodology makes to enable the soft attribute ranking problem to be addressed.

\subsection{Analysis}
\vspace*{-0.25\baselineskip}

Given that the fully supervised SWD algorithm significantly outperforms the unsupervised and weakly supervised methods on the Soft Attribute collection, we now analyze this model further by asking: How much rater training data is required to effectively learn a ranking function for a soft attribute (Sec.~\ref{sec:results:dataeff})? And, how does model performance change across different soft attributes (Sec.~\ref{sec:results:attrs})?

\begin{figure}[t]
  \includegraphics[width=.9\linewidth]{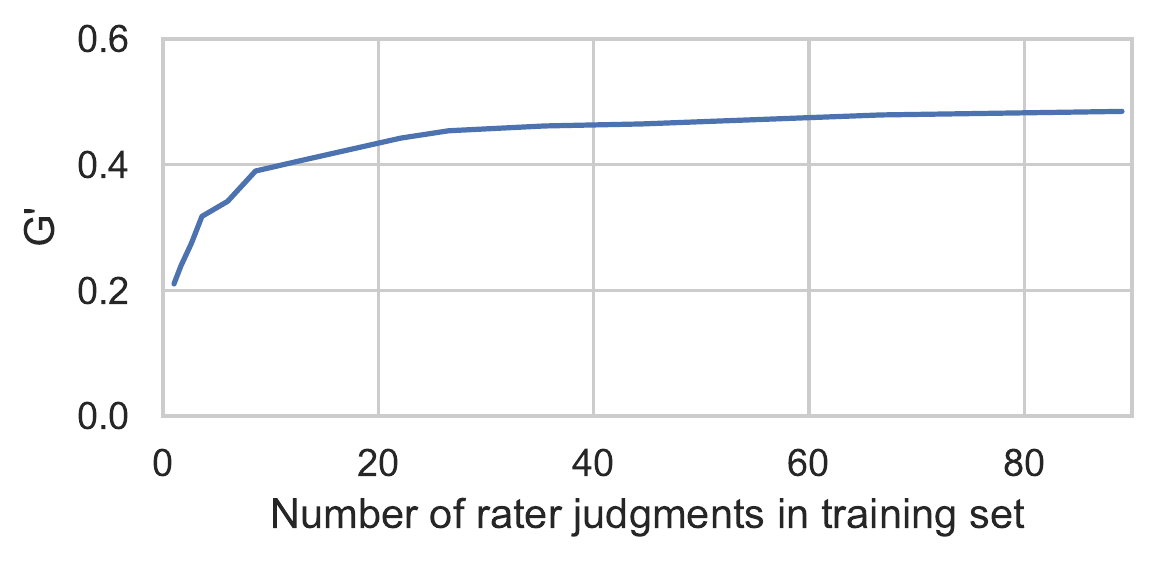}
  \vspace*{-0.75\baselineskip}
  \caption{SWD model performance (all soft attributes averaged) vs. the number of rater's judgments used for training.}
  \label{fig:learning_curve}
  \vspace*{-0.5\baselineskip}
\end{figure}

\vspace*{-0.25\baselineskip}
\subsubsection{Data Efficiency}
\label{sec:results:dataeff}

Figure \ref{fig:learning_curve} shows average SWD performance as a function of the number of rater's judgment (sets) provided to the learning algorithm. Specifically, we subsample raters in the training set and compute mean G', averaging over 25 runs at each subsample size. We see that the model is very data efficient: For any given soft attribute, judgments from approx. 20 raters are required to obtain near-optimal performance. This reinforces the value of pairwise preferences over a controlled sample of known items, as opposed to relying on binary judgments of tag presence or absence. 

\subsubsection{Performance Analysis by Soft Attribute}
\label{sec:results:attrs}

\begin{figure}[t]
  \vspace*{-0.25\baselineskip}
  \includegraphics[width=.75\linewidth]{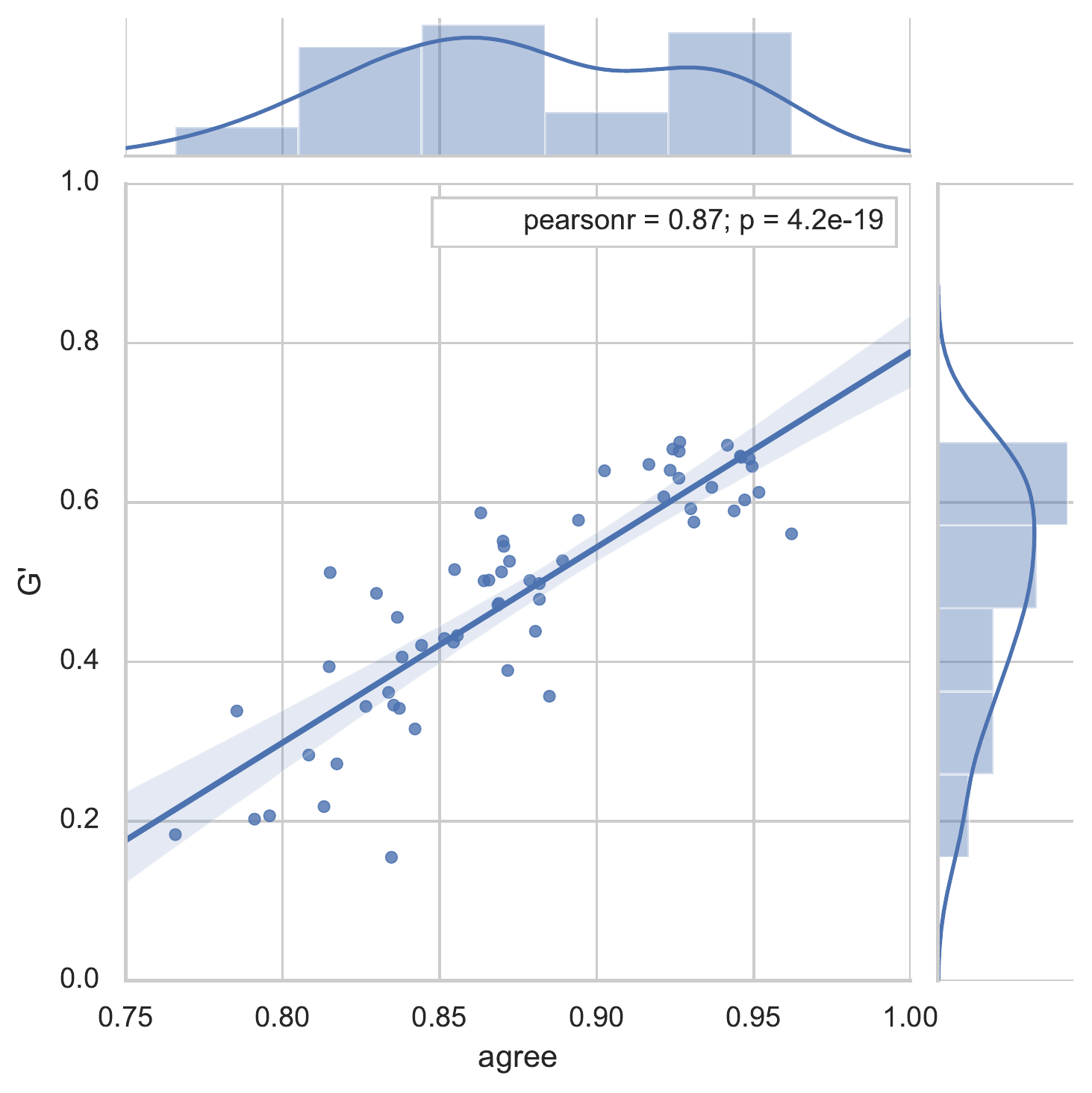}
  \vspace*{-0.75\baselineskip}
  \caption{SWD model performance (G') against attribute subjectivity (agree). Each point represents one soft attribute.}
  \label{fig:agree_vs_Gprime}
  \vspace*{-0.75\baselineskip}
\end{figure}

Next, we plot model performance against attribute subjectivity in Figure~\ref{fig:agree_vs_Gprime}.  We observe a clear correlation between the subjectiveness of a soft attribute (measured in terms of inter-rater agreement) and ranking performance (in terms of weighted gamma rank correlation $G'$).  As may be expected, soft attributes with less agreement are harder to predict, leaving significant room for personalized soft attribute scoring models as future work.
These results also suggest that predicting the ``softness'' of a soft attribute is also an important future direction.

%% file: 07-conclusions.tex
\vspace*{-0.75\baselineskip}
\section{Conclusions}

In this work, we have formalized recommender system critiquing based on \emph{soft attributes}, that is, aspects of items that cannot be established as universally agreed facts.
Developing a general methodology for obtaining soft attribute judgments, we have presented a dataset of pairwise preferences over soft attributes applied to the domain of movies. Our analysis of these preferences has shown that different attributes exhibit different levels of ``softness,'' with those that have less inter-user agreement being more difficult to model.
We have also developed approaches for critiquing based on soft attributes, in particular introducing a set of scoring approaches, from unsupervised to weakly supervised to fully supervised, employing both information retrieval approaches and embedding models. 
Moreover, comparisons on a standard tagging dataset has demonstrated how an appropriate data collection approach is key to making progress on the soft attribute scoring task.

Personalization is a key aspect of soft-attribute ranking that was not addressed in this paper, despite a clear signal of its importance. In fact, our analysis of performance by soft attribute suggests significant headroom for personalized soft attribute ranking methods as future work: The results suggest that ``softer'' soft attributes perform more poorly with non-personalized algorithms.